# Multipath Routing of Fragmented Data Transfer in a Smart Grid Environment

*Tuhin Borgohain
Department of Instrumentation Engineering, Assam Engineering College, Guwahati, India
borgohain.tuhin@gmail.com

Amardeep Borgohain
Department of Electrical Engineering, Assam Engineering College, Guwahati, India
amardeepborgohain@gmail.com

Rajdeep Borgohain
Department of Computer Science and Engineering, Dibrugarh University Institute of Engineering and Technology, Dibrugarh, India
rajdeepgohain@gmail.com

Sugata Sanyal
Corporate Technology Office
Tata Consultancy Services, Mumbai, INDIA
sugata.sanyal@tcs.com
*Corresponding Author

## ABSTRACT

The purpose of this paper is to do a general survey on the existing communication modes inside a smart grid, the existing security loopholes and their countermeasures. Then we suggest a detailed countermeasure, building upon the Jigsaw based secure data transfer [8] for enhanced security of the data flow inside the communication system of a smart grid. The paper has been written without the consideration of any factor of inoperability between the various security techniques inside a smart grid

## 1. INTRODUCTION

Scientific advances in the field of power generation and distribution has led to the development of the smart grid (SG). A typical smart grid facilitates the efficient distribution and automated metering of electricity amongst the end users. An SG employs a two way system of exchange of electricity and information for sustaining the whole power grid with minimal human interaction. This two way system can briefly be summarized as follows.

**Two way flow of electricity:** It refers to bidirectional flow of electricity between the smart grid and the users, the latter contributing by means of supplying electricity generated through commercial solar panels back into the grid [1].

**Two way flow of information:** A high performance communication network is necessary for automation and flow of information in an effective manner. The communication architecture consists of Neighbourhood Area Network (NAN), Building Area Network (BAN) and Home Area Network (HAN). NAN contains a number of BANs and each BAN is composed of a number of HANs. In HAN, the house hold appliances are connected to smart meters through wireless network like Zigbee, WiFi, Power line based communication. BAN collects information from HAN and delivers it to NAN. NAN gathers information from various units of BAN and sends it to the control centre. Smart meters are finally connected to a metering gateway. Smart meter is a device which delegates two way communications between the various communication units.

This inherent capability of automated functioning of the smart grid is made possible by a constant stream of information exchange over the internet (Zigbee, WiMAX, 3G), wired or wireless Ethernet, Bluetooth, fiber optics etc. [13]. But as in a desktop or a mobile computing device connected to the internet, the relay of data to and fro between the various micro level of the communication architecture of a SG (a detailed description of the communication framework can be found in [2]) is susceptible to cyber-attacks [26] as well as loss of valuable information by both the power station and customers [25]. As such the use of intrusion detection systems (IDS) and network security takes a paramount importance for safe confinement of data in their appropriate place in the grid.

## 2. OVERVIEW

In Section 3 and 4, we will analyze the different communication methods and the security loopholes in the existing smart grid models respectively. In Section 5, a brief analysis of the existing network security measures and IDS will be covered. In section 6, we will structure the whole multi-path routing of fragmented data transfer security system. In Section 7, we will conclude our paper with a brief description regarding the direction in which further research can be followed for improving the existing security and privacy status of the smart grid.

## 3. COMMUNICATION METHODS OF A SMART GRID

The various modes of communication can be categorized on two different basis viz.:
On the basis of mobility [1], [24]:

a. Wireless Technologies:
    i. Wireless Mesh Network (WMN)
    ii. Cellular Communication System

  iii. Cognitive Radio
  iv. IEEE 802.15
  v. Satellite Communications
  vi. Free Space Optical Communications

b. Wired Technologies
 i. Fiber Optic Communication
 ii. Power Line Communication

On the basis of communication requirements [3]:
a. Real Time Operational Communication Requirements
b. Administrative Operational Communication Requirements
c. Administrative Communication Requirements

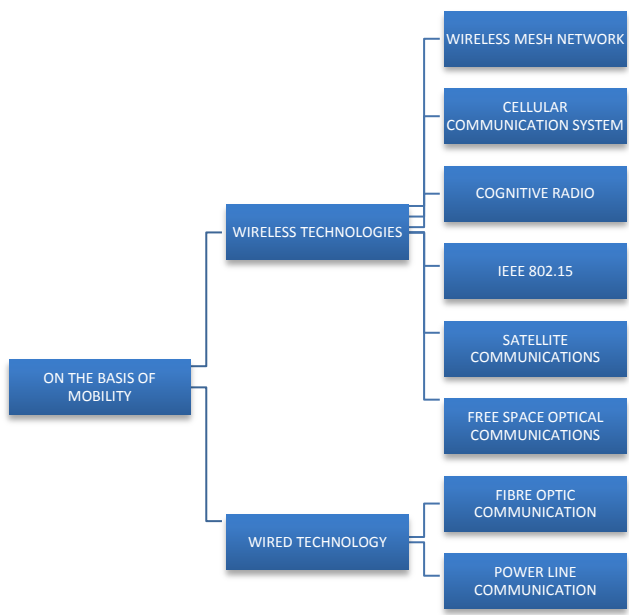

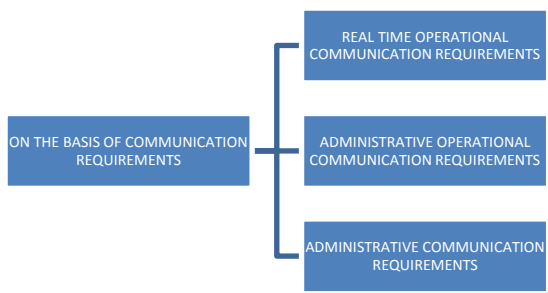

**Fig 1: Classification of communication modes in a smart grid**

## 4. SECURITY LOOPHOLES IN INTELLIGRID

A. *Malware spreading*
 Smart meters mostly communicate wirelessly. Spreading a malware into one of the meters can affect all other meters as these meters are organized in a mesh connected network.

B. *Reverse engineering*
 Tampering with the smart meters could possibly let the attackers turn someone else's power on and off or manipulate consumer's power bills and even allow the attackers to have easy access into the network to steal information and stage attacks on the grid. Attackers can tamper with the meters to steal power and can increase demand for power and vice-versa and even cause blackout.

C. *Encryption Key Vulnerability*
 Insecure implementation of security protocol having cryptographic encryption keys [23] on disk, hard-coded in software, or keeping in memory for extended periods of time could lead to security breach.

D. *DoS attack*
 Malicious threats such as Botnet can cause serious damage to the smart grid. A bot is an application that performs automated task over the internet. Number of Bots together is referred to as Botnet. Bot attacks involve threats like Denial-of-Service (DoS) attack that causes excess traffic in data transmission between nodes by circulating false information into the communication network resulting in delay or blockage in legitimate data transmission process.

E. *Stuxnet*
 Malware like Stuxnet [31] is basically designed to attack industrial control system specifically targeting programmable logic controllers (PLCs), intelligent electronic devices (IED), supervisory control and data acquisition (SCADA) which is widely used in industries and uses four zero-day flaws to attack machines using Microsoft Windows operating system. It is introduced using infected USB drives, weak passwords etc. The virus spreads into the network scanning for PLCs to inject its rootkit module to overwrite the legitimate commands and gain control over the grid system. Multiple rootkits are responsible for hiding the malicious codes from the operators to prevent detection of the Stuxnet.

## 5. EXISTING NETWORK SECURITY MEASURES

1. Security patches [17] mitigate any vulnerabilities or bugs in the grid network by updating the control system software like SCADA to improve the grid's performance. Security patches thwart malicious threats from exploiting the vulnerability in a grid system
2. Intrusion detection and protection system (IDPS) should be improvised
3. Bots in a network can be detected and removed using Standalone algorithm and network algorithm. [11]
4. Secure communications in cognitive radio [4].
5. Defence-in-depth is a network security measure for prevention of vulnerabilities into the control system (SCADA). It is a multi-layered approach which enables an operator to respond against cyber-attacks and consequently prevent security breach in the grid. The design is basically a series of concentric circles [16]. The first layer consists of firewalls along with demilitarized zone (DMZ) which creates a kind of sub network that acts as an additional layer to the defence of depth. A legitimate user can have access to information but an attacker cannot infiltrate into the SCADA system as the attacker would be stopped by the DMZ. The second layer consists of Public Key Infrastructure (PKI), key management, patch management. Key management has great importance in information security [15]. Authentication is the key to access control and having access to authentic public keys leads to authenticity for communication. The third layer consists of intrusion detection system (IDS) which monitor's network traffic. Any unusual activity in the network is typically informed to the operator. The network level consists of incident management [14] which involves in monitoring and responding to network intrusion. Defence of depth improvises itself by discovering new vulnerabilities.

6. Multifactor security authentication system [5] during access to the operational core of the smart grid for manual configuration.

7. Implementing cryptographic algorithms in conjunction with stenographic methods to make the information exchange between the various sectors of a smart grid invisible to third party unauthorized personnel. This can be facilitated by the implementation of data hiding techniques using prime numbers [6] and natural numbers [9].

8. Security threats such as malware, data theft can be minimized using supply chain cyber security [12] which augments the cyber security of the grid network within the supply chain.

9. The data transfer by use of IEEE 802.15 is susceptible to disruption in its continuous communication flow due to intrusion of malicious nodes in its mobile ad-hoc networks. These unwanted nodes can be detected and filtered from the required information flow by using promiscuous mode and appending NTNH field with RREP packet [7].

10. For continuous flow of information in the grid network which is the most challenging task in mobile ad-hoc network (MANET), ant colony optimization (ACO) [29] can be implemented.

11. For a secure communication over optical fiber, we can implement methods described in [10].

Moreover, many other network security measures like High Assurance Smart Grid [14], anonymizing smart grid metering data [27] can be implemented.

## 6. MULTI-PATH ROUTING OF FRAGMENTED DATA TRANSFER:

Implementing two computing devices at the sending and receiving terminal inside a smart grid, we can secure the interexchange of information by the use of an automated sliced multi-routing jigsaw based secure data transfer [8] with minimal manual interaction. The two computing devices are kept in sync using a secure cloud based technology as in [18], [19], [20], [21], [22], [28] and [30].

In case of the automated sliced multi-routing Jigsaw based data transfer the following steps takes place sequentially:

i. A packet of data X is divided into N-parts by the sending computing device (A) and each part is assigned a sequence number based on its data content $(X_1, X_2,...., X_N)$.

ii. The sequence numbers (say w1) are synced with the receiving computer terminal (B) over the cloud and then the N-parts are shuffled.

iii. Each part of the data packet is then transformed to X' by performing the operation $1Xi1$ (XOR) Pi for all "i" ranging from 1 to k-1, followed by performing transform (P, R) which generates a new random number and this number is assigned to R. ([8]).

iv. Using a unique encryption key, the new random numbers are shuffled in a random manner and its decryption key (w2) is computed at (A) and then synced to (B).

v. The shuffled random numbers are then sliced into M-equal parts, M being constant the value of which is then synced to (B).

vi. Before slicing of each part of the data packet into M equal parts, (A) assigns a sub-sequence number (say w3) to each slice of the random numbers which is again synced with (B).

vii. The sliced M parts of the random numbers are then sent to (B) through M different routes. In each route only one slice of each random number is allowed and based on the divisibility of M by the lowest factor (Say M is divisible by v), the slices are sent in random bundles of "v" slices together. This factor "v" is synced with (B). In this way, any third party interference would only result in obtaining an unintelligible sub-part of a part of a single data packet which is different from the true form of the data packet.

viii. Upon reaching (B), the slices unbundle themselves independently and the M different slices of each random numbers are first matched and formed into N random numbers using the synced sub-sequence numbers (w3) of each random number.

ix. After the sub-sequence matching, the random numbers are reshuffled to their original positions using the previously synced decryption key (w2) and the whole packet is obtained albeit in the form of combination of random numbers derived through the process in [8].

x. The Message Authentication Code (MAC) (for full details refer [11]) for each packet of data is calculated and then compared to the MAC previously assigned to each packet and synced with (B). This is performed as a check for any tampering of data to alert the whole system.

xi. The random numbers are converted back to their original form of the part of each data pack (again refer to [8] for the details of the conversion process).

xii. After conversion of the random numbers to their original form, they are shuffled back into their original position by matching of their sequence numbers (w1) in accordance to their position in the data.

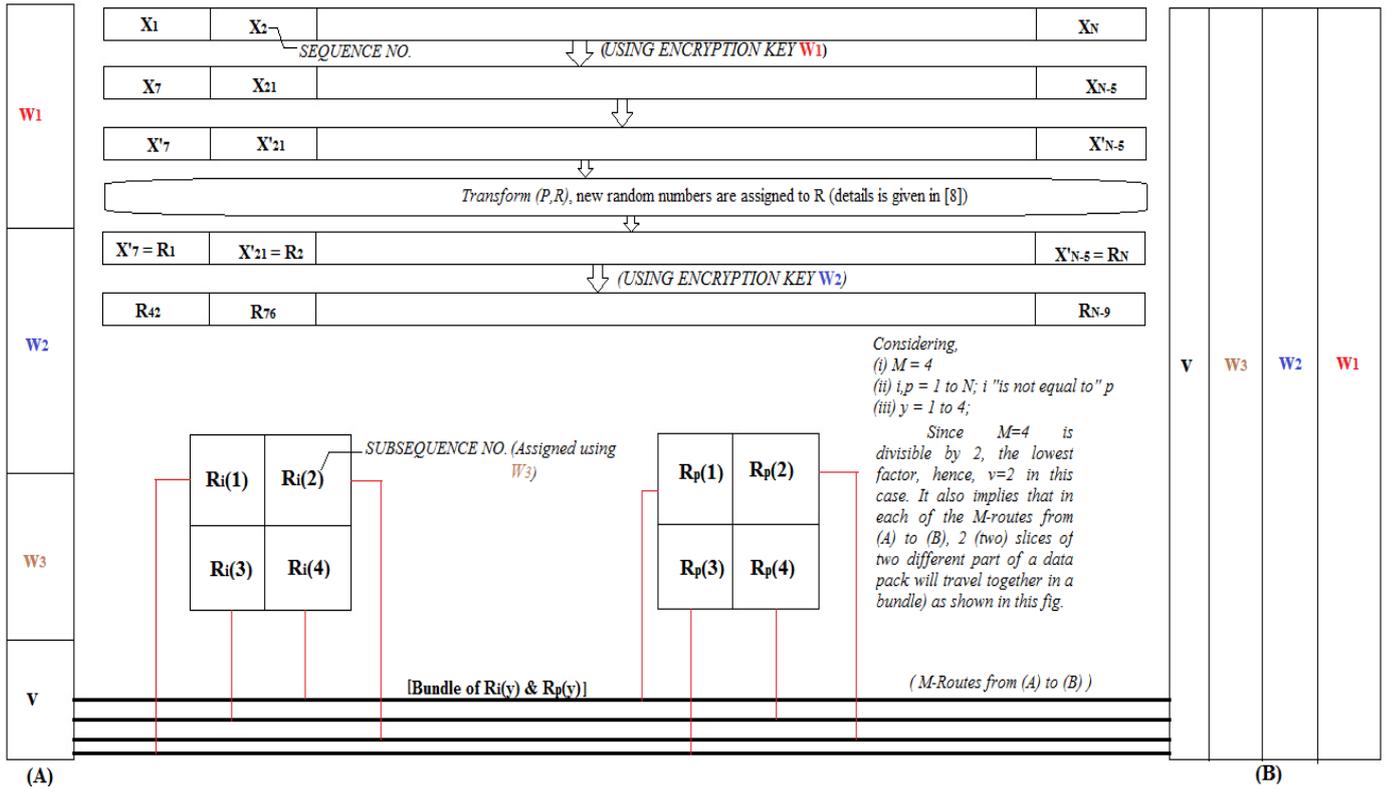

Fig 2: Operational flow diagram of Multi-Path Routing of Fragmented Data Transfer

## 7. CONCLUSION

We achieved our objective of this paper by centralizing all the various communication modes inside a smart grid, analyzed their drawbacks from the perspective of security and privacy along with a general survey on the existing security measures addressing the various security loopholes. Then we proposed a security measure for enhanced security by building upon the Jigsaw based secure data transfer method. In our proposed method, we have minimized the stages of human interaction inside the communication system of a smart grid to keep the core communication layout inaccessible to third party intruders. As a conclusion to our paper, we would acknowledge the fact that a boost in the cloud computing encryption techniques and implementation of additional encryption and stenographic techniques to the bundling of the sliced parts of the data would prove to be much fruitful in enhancing the security of the multi-path routing of fragmented data transfer to a whole new level